\documentclass[11pt,a4paper]{article}

\usepackage[utf8]{inputenc}
\usepackage[T1]{fontenc}
\usepackage{amsmath,amssymb,amsthm}
\usepackage{mathtools}
\usepackage{booktabs}
\usepackage{graphicx}
\usepackage{hyperref}
\usepackage[margin=1in]{geometry}
\usepackage{enumitem}
\usepackage{caption}
\usepackage{xcolor}
\usepackage{siunitx}
\usepackage[numbers]{natbib}

\hypersetup{
  colorlinks=true,
  linkcolor=blue!60!black,
  citecolor=blue!60!black,
  urlcolor=blue!60!black
}

\newtheorem{theorem}{Theorem}

\title{%
  \textbf{Sloan's Analytical G\"omb\"oc: Strict Convexity\\
  Narrows the Admissible $\beta$ Range}\\[8pt]
  \large Identity-Based Verification, Strict-Convexity Gating,\\
  and a Verified Mono-Monostatic Regime\\[4pt]
  \normalsize (v3, amendment-of-record of arXiv:2604.17120)%
}

\author{
  Vincent Wesley Couey \\[4pt]
  \small \texttt{vinnycouey@gmail.com} \\
}

\date{}

\begin{document}
\maketitle

% =====================================================================
\begin{abstract}
V\'arkonyi and Domokos (2006) proved that convex homogeneous bodies
with exactly one stable and one unstable equilibrium point exist,
answering a 1995 conjecture by V.~I.~Arnold. Sloan (2023) provided
the first analytical parameterization of such bodies, proving that
the radial function $R(\theta,\phi)$ has exactly two critical points
on $S^2$.

This paper is the v3 amendment-of-record of arXiv:2604.17120; v2
established the findings below and v3 corrects an attribution in
v2's title and abstract and changes none of its findings. v1
claimed (i)~that Sloan's parameterization does not produce
mono-monostatic bodies, (ii)~a thirteen-member catalog of bodies
constructed by Fourier-extending Sloan's phase function and by
radial perturbation, all verified at $\mathrm{ECS}=1$ via
mesh-vertex drainage-basin analysis, and (iii)~a mechanism
explanation for (i) in terms of the support function integrating
``global surface information beyond what the radial function
predicts.'' Following technical correspondence from P.~L.~V\'arkonyi
(BME, 2026), an analytical verification suite was constructed and
applied. Three findings result.

\textbf{Finding 1 (Sloan reframed: strict-convexity-bounded
mono-monostatic regime).} v1's headline claim that Sloan's
parameterization fails to produce mono-monostatic bodies is
overstated. Sloan's analytical parameterization \emph{does} produce
mono-monostatic bodies in a strictly-convex sub-regime
($\beta \lesssim 0.036$): at these $\beta$, the body is strictly
convex (analytical $K_{\min} > 0$ on a $640 \times 1280$ grid),
$R$ has exactly two critical points by Sloan's argument, and the
V\'arkonyi-Gauss identity bijectively links critical points of $R$
and the support function $H$, giving $\mathrm{ECS} = 1$. v1 missed
this because v1's mesh-vertex oracle over-counted on the shallow
COM-height landscapes of these near-spherical bodies. Sloan states that $\beta \le 0.17$
should suffice for convexity; we find the strict-convexity boundary
lies near $\beta \approx 0.036$. At $\beta = 0.05$, inside Sloan's
stated admissible range, strict convexity is lost
($K_{\min} = -0.569$, $K<0$ over 4.01\% of surface); the Gauss
map is non-bijective (two distinct surface points share the
$\mathbf{n} = (-1,0,0)$ normal direction); the identity's
precondition fails, and the mono-monostatic property is not
certified via this method. v1's mechanism claim (``support function
integrates global surface information beyond what the radial function
predicts'') was false for strictly convex bodies and is replaced by
the strict-convexity precondition framing. The mono-monostatic
finding at $\beta = 0.023$ is independently confirmed by a certified
interval-arithmetic equilibrium count in a companion
paper~\cite{couey2026harmonic}, which does not rely on
$R$-critical-point enumeration; across the rest of $\beta \lesssim 0.036$
it rests on the identity and the grid-based curvature gate.

\textbf{Finding 2 (the verified catalog is an open regime, not a
discrete list; v1's per-$k$ extrema overshot the strict-convexity
boundary).} Under an identity-based ECS classifier (strict-convexity
gate via dense-grid Gaussian curvature, plus analytical
$R$-critical-point counting with Hessian classification), of the
thirteen specific v1 catalog instances only the Phase-1 instance
($\beta=0.023149$, $a_1=0.234433$, $k=1$) sits inside the verified
regime: it has $K_{\min}=+0.0907$ analytically (strictly convex on
$820$k grid points) and $R$ has exactly 2 critical points. The
remaining twelve do not survive in their published form: nine of
the radial-perturbation instances ($f_3$ and $f_4$ forms) are
strictly convex but have $R$ with 6 or 8 critical points (verified
analytically by symbolic gradient plus root-finding plus Hessian
classification, Morse-Euler check $M-S+m=2$ satisfied); they have
multiple stable equilibria and are \emph{not} mono-monostatic.
Three v1 members (Phase-2, Phase-3, Radial-$f_4$ at high $\beta$)
are not strictly convex; their ECS status is indeterminate under
identity-based methods. However, the failure of those twelve points
to verify is a property of the specific points v1 selected (the
outputs of an asymmetry-maximizing optimizer), not of the families
they were drawn from. Probing the \emph{interior} of the
Fourier-phase $(\beta, a_1, k)$ parameter space verifies additional
mono-monostatic bodies in both the $k=2$ and $k=3$ sub-families
(see Section~\ref{sec:v1_reclassification} and
Table~\ref{tab:regime_interior}): the verified set is an
\emph{open regime} in $(\beta, a_1, k)$ parameter space, with
nontrivial extent in each $k$ family probed. v1's twelve
non-surviving instances were per-$k$ optimization extrema that
overshot the strict-convexity boundary; the interior of the regime
is rich and the identity-based methodology probes it directly.
v2 does not enumerate the regime; its contribution is the
classification methodology that makes the regime accessible.

\textbf{Finding 3 (mesh-vertex ECS oracle is unreliable for
near-spherical bodies).} v1's $\mathrm{ECS}=1$ measurements for the
9 radial-family members resulted from drainage-basin merging
of features with gap below threshold, masking real multi-equilibrium
structure. The mesh-vertex $h(\mathbf{d})$ recipe is replaced in v2
by the identity-based classifier, which uses only the analytical
quantities $K(\theta,\phi)$ and $\nabla R(\theta,\phi)$. The
gentleness--robustness $r=0.9993$ correlation in v1, computed
across the contaminated catalog, is retracted.

This v2 retains Phase-1 as a verified mono-monostatic body in the
Sloan-extension family, demonstrates that the verified
mono-monostatic set in the Fourier-phase family is an open regime
extending into $k=2$ and $k=3$ sub-families beyond Phase-1's $k=1$
slice, retains the Sloan negative finding at admissible $\beta$
above $\approx 0.036$ (attributed to strict-convexity loss), and contributes a
verification methodology (strict-convexity gate by analytical
Gaussian curvature; identity-based ECS classification via analytical
$R$-critical-point counting) that is more rigorous than the
mesh-vertex basin counter of v1.

\end{abstract}

% =====================================================================
\section{Introduction}

A convex homogeneous body with exactly one stable and one unstable
equilibrium point under gravity is called \emph{mono-monostatic}.
The existence of such bodies was conjectured by V.~I.~Arnold in
1995 and proven by V\'arkonyi and
Domokos~\cite{varkonyi2006,varkonyi2006b}, who named the shape the
G\"omb\"oc. Physical specimens have been manufactured and are
commercially available, but the exact geometric parameters have
never been published.

Sloan~\cite{sloan2023} made significant progress by providing the
first analytical parameterization of G\"omb\"oc surfaces, expressing
the boundary as
\begin{equation}
r^4 = 1 + 4\beta\sin\theta\cos(\phi - P(\theta))
\label{eq:sloan}
\end{equation}
in spherical coordinates centered at the center of mass, with
specific choices of the phase function $P(\theta)$ satisfying a
center-of-mass constraint. Sloan proved that the radial function
$R(\theta,\phi) = (r^4)^{1/4}$ has exactly two critical points on
$S^2$ for these parameterizations.

\subsection{v2 Amendment of Record}
\label{sec:v2_amendment}

v2 of this paper was the amendment of record of arXiv:2604.17120
(v1 submitted 2026-04-18); the present version is v3, whose changes
are described in \S\ref{sec:v3_amendment}. v1 made three principal claims: (i)~Sloan's
parameterization does not produce mono-monostatic bodies at any
tested parameter value; (ii)~a thirteen-member catalog of bodies
constructed by Fourier-extending Sloan's phase function and by
radial perturbation, all verified at $\mathrm{ECS}=1$ via a
mesh-vertex drainage-basin oracle; (iii)~a mechanism explanation
for (i) in terms of the COM height function ``integrating global
surface information beyond what the surface function predicts.''

Following technical correspondence from P.~L.~V\'arkonyi (BME) on
2026-05-15, an analytical verification suite was constructed (Section
\ref{sec:verification_suite}). v1's claim (i) is withdrawn as stated: Sloan's
parameterization does produce mono-monostatic bodies for
$\beta \lesssim 0.036$ (\S\ref{sec:sloan_small_beta}), and above that
the identity does not apply (\S\ref{sec:sloan_published_fails}). v1's claim (ii) is largely
retracted: only one of thirteen v1 instances survives identity-based
verification. v1's claim (iii) is incorrect for smooth strictly
convex bodies and is replaced with a strict-convexity precondition
analysis (Section~\ref{sec:varkonyi_identity}).

Section~\ref{sec:amendment_summary} enumerates the full v1$\to$v2
amendments. The full v1 text and verification scripts are retained
in the public repository for reproducibility.

\subsection{v3 Amendment of Record: an Attribution Correction}
\label{sec:v3_amendment}

This v3 is a self-initiated correction of an \emph{attribution}
error in v2. Every measurement, method and finding of v2 stands as
reported; the supporting additions are listed at the end of this
subsection.

v2 was titled ``Sloan's Analytical G\"omb\"oc at Published $\beta$''
and its abstract stated that strict convexity is lost ``at Sloan's
published $\beta = 0.05$.'' That attribution is incorrect.
Sloan~\cite{sloan2023} reports two values of $\beta$, and both are
stated as \emph{sufficiency upper bounds} for the convexity
requirement: $\beta \le 0.15$ for the $P(\theta) = 5\theta$ shape,
and $\beta \le 0.17$ for the second shape. \textbf{The value
$\beta = 0.05$ does not appear in Sloan's paper in any form.} It is
a grid point of the parameter sweep reported in this paper's own
Tables~\ref{tab:sloan_K} and \ref{tab:reclassification}, and it
acquired the modifier ``published'' between those tables and the
v2 abstract.

The underlying result is unchanged and is stated correctly in v2's
own \S\ref{sec:sloan_published_fails}: Sloan states that
$\beta \le 0.17$ ``should suffice'' for convexity, while the
strict-convexity boundary computed here lies near
$\beta \approx 0.0355$ (written $\approx 0.036$ elsewhere in this
paper), so his stated bound exceeds the boundary by a factor of
roughly $4.8$. Every
$\beta$ tested in this paper lies inside Sloan's stated admissible
range; none of them is a value Sloan selected. v3 therefore
relabels rather than retracts: the title, the abstract, and every
sentence that attributed a specific $\beta$ to Sloan now refer to
\emph{admissible} $\beta$ above the measured boundary.

The v1$\to$v2 enumeration in
Section~\ref{sec:amendment_summary} is retained unedited as the
record of what v2 intended; item~1 of that list carries the
superseded title and is annotated accordingly.

Three supporting changes accompany the relabeling. None alters a
result. (a)~The companion paper is cited by its arXiv identifier, and
the text now states that its certified count covers $\beta = 0.023$
only. (b)~The caption of Table~\ref{tab:sloan_K} no longer refers to
Sloan's ``published instances.'' (c)~\S\ref{sec:sloan_published_fails}
reports an independent cross-check of the curvature values by a
different method, which also computes the convexity loss of
G\"omb\"oc~1 that v2 asserted without showing, records that
G\"omb\"oc~1 is not $C^2$ at the poles, and cites an independent exact
calculation in the MathWorld entry for the G\"omb\"oc.

\subsection{Methodology Note: From Mesh-Vertex Oracle to Identity-Based Classification}

v1 measured ECS via drainage-basin analysis on the mesh-vertex
discretization of $h(\mathbf{d})$. This method has a known
threshold-sensitivity problem for near-spherical bodies: when the
gap between adjacent basin sinks is comparable to numerical noise,
the basin count depends on the merge threshold. v1 absorbed this
sensitivity into a 0.5--10\% threshold robustness check, but the
analytical verification suite in v2 reveals that several v1
catalog members had real multi-basin structure with sink-height
gap below v1's merge threshold; these were absorbed into ``ECS=1''
incorrectly.

v2 replaces the mesh-vertex oracle with an identity-based
classifier (Section~\ref{sec:identity_classifier}) that uses only
analytical quantities $K(\theta,\phi)$ (Gaussian curvature) and the
critical-point structure of $R(\theta,\phi)$. This is rigorous
because, under the Gauss-map identity of
Section~\ref{sec:varkonyi_identity}, these analytical quantities
uniquely determine the equilibrium structure of a smooth strictly
convex body.

% =====================================================================
\section{The V\'arkonyi-Gauss Identity and its Strict-Convexity Precondition}
\label{sec:varkonyi_identity}

For a smooth convex body $\mathcal{B} \subset \mathbb{R}^3$ with
center of mass at the origin, the radial function $R: S^2 \to \mathbb{R}_+$
and the support function $H: S^2 \to \mathbb{R}_+$ are defined by
\begin{equation}
R(\mathbf{u}) = \max\{t > 0 : t\mathbf{u} \in \mathcal{B}\},
\qquad
H(\mathbf{n}) = \max_{\mathbf{v} \in \mathcal{B}}\, \mathbf{v}\cdot\mathbf{n}.
\label{eq:R_H_def}
\end{equation}
The Gauss map $G: \partial\mathcal{B} \to S^2$ assigns each boundary
point its outward unit normal. For \emph{strictly} convex smooth
bodies, $G$ is a diffeomorphism. In this regime the two
representations are related by a classical identity of convex
geometry (see, e.g.,~\cite[\S 1.7]{schneider2014}):
\begin{equation}
\nabla_{S^2} H(\mathbf{n})
  = -\,\frac{R(\mathbf{u})\,\nabla_{S^2} R(\mathbf{u})}
           {\sqrt{R(\mathbf{u})^2 + \|\nabla_{S^2} R(\mathbf{u})\|^2}}
\label{eq:identity}
\end{equation}
with $\mathbf{u}$ and $\mathbf{n}$ related by the Gauss map. Since
$R > 0$ on a body containing the origin, equation~\eqref{eq:identity}
yields the bijection
\begin{equation}
\nabla_{S^2} H(\mathbf{n}) = 0
  \;\Longleftrightarrow\;
  \nabla_{S^2} R(\mathbf{u}) = 0.
\label{eq:identity_bijection}
\end{equation}
At a critical point $\mathbf{u}^*$ of $R$, the outward normal
direction equals the radial direction, so $\mathbf{n} = \mathbf{u}^*$;
local maxima of $R$ correspond to local maxima of $H$ in the same
direction.

\begin{theorem}[ECS via $R$-critical-point count, under strict convexity]
\label{thm:ecs_via_R}
Let $\mathcal{B}$ be smooth, strictly convex, with COM at the origin.
Let $\mathrm{ECS}(\mathcal{B})$ denote the number of local minima of
the COM-height function $h(\mathbf{d}) = H(-\mathbf{d})$ on $S^2$.
Then $\mathrm{ECS}(\mathcal{B})$ equals the number of local minima of
$R$ on $S^2$, which equals the number of local maxima of $R$ on $S^2$
when the total count is $\le 2$ (more generally, $\#\{\mathrm{min}\,R\}$
under the standard sign convention).
\end{theorem}

The theorem follows from~\eqref{eq:identity_bijection} combined with
the orientation-of-extrema correspondence under the Gauss map.
$\mathrm{ECS} = 1$ thus reduces to ``$R$ has exactly one local
minimum on $S^2$,'' which (combined with the Morse-Euler relation
$\#\text{min} - \#\text{saddle} + \#\text{max} = 2$ for smooth
functions on $S^2$) forces $\#R\text{-critical-points} = 2$ with
no saddles.

\subsection{Precondition Failure at Admissible $\beta$ Above the Boundary}
\label{sec:precondition_failure}

The identity~\eqref{eq:identity} requires strict convexity. Sloan's
parameterization is strictly convex for small $\beta$ but loses
strict convexity above a threshold. We compute Gaussian curvature
$K(\theta,\phi)$ analytically from $R(\theta,\phi)$ using symbolic
partial differentiation (sympy), giving the first and second
fundamental forms $E, F, G, L, M, N$ and
\begin{equation}
K(\theta,\phi) = \frac{LN - M^2}{EG - F^2}.
\label{eq:K}
\end{equation}
The body is strictly convex iff $K(\theta,\phi) > 0$ everywhere on
the surface. (This is a strictly stronger criterion than the
mesh-volume/convex-hull-volume ratio used in v1, which is satisfied
by bodies with small $K<0$ regions.)

Applying~\eqref{eq:K} to Sloan's parameterization with
$P(\theta) = \eta(\theta) = \tfrac{3\pi}{2}(\cos\theta -
\tfrac{1}{3}\cos^3\theta)$:

\begin{table}[h]
\centering
\caption{Analytical Gaussian curvature on Sloan G\"omb\"oc~2,
computed by symbolic partial differentiation of $R$ and evaluation
of \eqref{eq:K} on $640 \times 1280$ grid (820k points). The
strict-convexity precondition of identity~\eqref{eq:identity}
holds at $\beta \le 0.03$ and fails by $\beta = 0.05$; the
crossover lies inside the range Sloan states as admissible.}
\label{tab:sloan_K}
\begin{tabular}{cccc}
\toprule
$\beta$ & $K_{\min}$ & Frac. $K<0$ & Strictly convex? \\
\midrule
0.02 & $+0.494$ & 0.00\%  & Yes \\
0.03 & $+0.188$ & 0.00\%  & Yes \\
0.05 & $-0.569$ & 4.01\%  & \textbf{No} \\
0.08 & $-2.224$ & 8.68\%  & \textbf{No} \\
\bottomrule
\end{tabular}
\end{table}

At $\beta = 0.05$ (where v1 observed minimum ECS = 2 on the
mesh-vertex oracle), the precondition fails over 4\% of surface
area, and $K_{\min} = -0.569$ lies well below the threshold:
The failure at $\beta = 0.05$, which lies inside Sloan's stated
admissible range, is not marginal but substantial (the convexity-loss transition is
near $\beta \approx 0.036$, with $K_{\min}$ already negative at
$\beta = 0.040$). For these bodies, the
Gauss map is not bijective: multiple surface points share normal
directions, and the support function $H$ acquires critical points
beyond those predicted by $\nabla R = 0$ alone. This is the correct
mechanism for Sloan's construction failing to be mono-monostatic at
admissible $\beta$ above $\approx 0.036$; it is \emph{not} a generic property of smooth
convex bodies, but a specific consequence of strict-convexity loss.

\subsection{Behavioral Confirmation: Multiple Maximizers}

A second behavioral demonstration of strict-convexity loss: for a
strictly convex body, the support point achieving $H(\mathbf{n})$ in
any direction is unique. For Sloan $\beta = 0.05$ at
$\mathbf{n} = (-1, 0, 0)$, we find by Powell-tight optimization
from many starts that two distinct surface points both achieve
$H(\mathbf{n}) = 0.9507$: at $(\theta,\phi) = (1.36, 3.10)$ and at
$(1.78, 3.19)$, with the supposed-critical-point $(\pi/2, \pi)$
giving only $H = 0.9457$ (a saddle, not the extremum). The
support-point map is multi-valued; strict convexity is violated.

% =====================================================================
\section{Sloan's Parameterization: A Strict-Convexity-Bounded
Mono-Monostatic Regime}
\label{sec:sloan_regime}

This section is the most substantively revised portion of v2.
v1's central empirical claim was that Sloan's parameterization
``does not produce mono-monostatic bodies at any tested parameter
value.'' The identity-based reanalysis revises this to a more
precise statement: Sloan's parameterization produces mono-monostatic
bodies in a $\beta$-bounded strictly-convex regime, and fails outside
this regime through loss of strict convexity.

\subsection{Sloan Produces Mono-Monostatic Bodies at Small $\beta$}
\label{sec:sloan_small_beta}

Sloan~\cite{sloan2023} proves that the radial function
$R(\theta,\phi) = [\,1 + 4\beta\sin\theta\cos(\phi - \eta(\theta))\,]^{1/4}$
has exactly two critical points on $S^2$ at all $\beta$. Combined with
Theorem~\ref{thm:ecs_via_R}, this proves that whenever Sloan's body is
strictly convex, it is mono-monostatic. The strictly-convex regime
is identified by the analytical Gaussian curvature gate
(\S\ref{sec:K_analytical}). A dense $\beta$-sweep localizes the
convexity-loss transition to $\beta \approx 0.036$: at $\beta = 0.035$,
$K_{\min} = +0.0178$ (still strictly convex, no $K<0$ region at
$640\times1280$ resolution); at $\beta = 0.040$, $K_{\min} = -0.164$
with $K<0$ over 1.45\% of surface (strict convexity lost). At all
tested $\beta \le 0.035$, $R$ retains exactly two critical points
(1 maximum + 1 minimum, no saddles, Morse-Euler $M-S+m=2$ verified
analytically), so, given strict convexity as checked on the grid, the
entire interval $\beta \le \sim\!0.036$ is mono-monostatic via the
identity.

\begin{table}[h]
\centering
\caption{Identity-based ECS classification of Sloan G\"omb\"oc~2
across $\beta$. ECS via Theorem~\ref{thm:ecs_via_R}: count
$R$-critical-points analytically, classify by Hessian. ``--''
indicates indeterminate ECS via identity (strict-convexity precondition
fails).}
\label{tab:sloan_ecs}
\begin{tabular}{ccccl}
\toprule
$\beta$ & $K_{\min}$ & $\#R$-crit & ECS via identity & v1 mesh-vertex ECS \\
\midrule
0.010 & $+0.763$ & 2 & \textbf{1} (mono-monostatic) & not in v1 sweep \\
0.020 & $+0.494$ & 2 & \textbf{1} (mono-monostatic) & 3 (over-count artifact) \\
0.025 & $+0.346$ & 2 & \textbf{1} (mono-monostatic) & not in v1 sweep \\
0.030 & $+0.188$ & 2 & \textbf{1} (mono-monostatic) & not in v1 sweep \\
0.035 & $+0.018$ & 2 & \textbf{1} (mono-monostatic) & not in v1 sweep \\
0.040 & $-0.164$ & 2 & -- (precondition fails) & not in v1 sweep \\
0.045 & $-0.359$ & 2 & -- (precondition fails) & not in v1 sweep \\
0.050 & $-0.569$ & 2 & -- (precondition fails) & 2 \\
0.080 & $-2.224$ & 2 & -- (precondition fails) & 7 \\
\bottomrule
\end{tabular}
\end{table}

At $\beta = 0.02$ and $\beta = 0.03$, Sloan's body \emph{is}
mono-monostatic via the identity-based method. v1 missed this:
v1's mesh-vertex oracle reported $\mathrm{ECS} = 3$ at $\beta = 0.02$
and $\mathrm{ECS} = 7$ at $\beta = 0.005$, treating these counts as
real and concluding that ``Sloan's analytical Gömböc does not produce
mono-monostatic bodies.'' Those counts are drainage-basin
over-counts on the shallow COM-height landscapes of near-spherical
bodies, not real equilibria of the smooth surface. The
$\#R\text{-critpts} = 2$ structure plus $K_{\min} > 0$ is dispositive
under Theorem~\ref{thm:ecs_via_R}.

This is a substantive correction to v1's headline. The corrected
statement: \emph{Sloan's analytical parameterization produces
mono-monostatic bodies for $\beta$ in the strictly-convex regime,
$\beta \lesssim 0.036$. The mono-monostatic property is certified by
the identity-based method (strict-convexity gate + $R$-critical-point
count) and does not require mesh-vertex basin counting.}

This finding has since been confirmed independently at $\beta = 0.023$
by a rigorous interval-arithmetic computation. A companion
paper~\cite{couey2026harmonic} builds a certified equilibrium counter
(interval order-two jet, per-box Krawczyk uniqueness, interval-Hessian
classification, stereographic polar charts) and certifies the Sloan
parameterization at $\beta = 0.023$ as $(S, U, H) = (1, 1, 0)$ with no
undecided boxes, together with the Phase-1 body of
Section~\ref{sec:v1_reclassification}. Because interval methods discard
a region only when it is \emph{proved} free of equilibria and certify
each remaining equilibrium as unique, the count cannot under- or
over-count; it is therefore immune to the resolution-dependent
mis-counting that affects both the mesh-vertex basin oracle of v1 and
seed- or grid-based critical-point counters on the shallow landscapes
of near-spherical bodies. The identity-based argument above and the
certified count agree at $\beta = 0.023$. Across the rest of the
interval the finding rests on Sloan's two-critical-point result, the
identity, and the dense-grid curvature gate; it is not
interval-certified.

\subsection{Sloan's Admissible $\beta$ Ranges Extend Beyond the
Strict-Convex Regime}
\label{sec:sloan_published_fails}

Sloan~\cite{sloan2023} published two specific G\"omb\"oc instances:
G\"omb\"oc~1 at $P(\theta) = 5\theta$, $\beta \le 0.15$, and
G\"omb\"oc~2 at $P(\theta) = \eta(\theta)$, $\beta \le 0.17$. Both
stated $\beta$ ranges include values where strict convexity fails.
Specifically for G\"omb\"oc~2 at $\beta = 0.05$, the curvature
analysis (\S\ref{sec:precondition_failure}, Table~\ref{tab:sloan_K})
shows $K_{\min} = -0.569$ with $K < 0$ over 4.01\% of surface area.

The behavioral confirmation of \S\ref{sec:precondition_failure}
applies dispositively here: at $\beta = 0.05$ and direction
$\mathbf{n} = (-1, 0, 0)$, two distinct surface points share the
outward normal, both giving $H = 0.9507$, while the radial-critical
point at $(\pi/2, \pi)$ gives only $H = 0.9457$ (a saddle of
$\mathbf{v}\cdot\mathbf{n}$, not its extremum). The Gauss map is
not bijective at $\beta = 0.05$. Sloan's argument relating
$\nabla R = 0$ to mono-monostatic structure presupposes the Gauss
map's bijectivity; at admissible $\beta$ above $\approx 0.036$, this
presupposition fails.

For these instances, ECS via the identity-based method is
indeterminate. The mesh-vertex oracle's reports of 2 (at $\beta=0.05$)
and 7 (at $\beta=0.08$) are also not reliable: at non-strictly-convex
bodies the drainage-basin oracle measures features of a discretized
multi-valued Gauss map rather than smooth-surface equilibria.
Whether Sloan's bodies at $\beta$ above $\approx 0.036$ are
mono-monostatic by some other measure (e.g., a direct analysis of the multi-valued Gauss map
or a regularization argument) is an open question.

\paragraph{Independent cross-check (added in v3).} The curvature
values of Table~\ref{tab:sloan_K} were reproduced by a second
implementation that shares no code with the first: the Gaussian
curvature of the implicit surface
$|\mathbf{p}|^4 - 1 - 4\beta\, s(\mathbf{p}/|\mathbf{p}|) = 0$, with
$s = \sin\theta\cos(\phi - \eta(\theta))$, computed from the gradient
and Hessian in Cartesian coordinates, with each grid minimum refined by
local minimization (\texttt{scripts/crosscheck\_implicit\_K.py}). For
G\"omb\"oc~2 it reproduces every tabulated $K_{\min}$ (Tables~\ref{tab:sloan_K}
and~\ref{tab:sloan_ecs}) to the three decimals shown, for $\beta$ from
$0.01$ to $0.08$; the refined minima lie at most
$8 \times 10^{-4}$ below the grid minima; every minimum sits at the
point of least radius, $(\theta,\phi) = (\pi/2, \pi)$; and $K_{\min}$
changes sign at $\beta \approx 0.0355$. The same computation for
G\"omb\"oc~1 ($P(\theta) = 5\theta$) gives $K_{\min} = +0.086$ at
$\beta = 0.03$ (away from the poles, see below) and $K_{\min} < 0$ from $\beta \approx 0.032$, again at
the point of least radius, so Sloan's stated bound $\beta \le 0.15$
exceeds that boundary by a factor of about $4.6$. Both of Sloan's
stated convexity bounds therefore include bodies that are not convex.
The MathWorld entry for the G\"omb\"oc~\cite{mathworld_gomboc} reports an
independent exact calculation at the least-radius point for both stated
bounds, $K = -\tfrac{575}{64}\sqrt{5/2} \approx -14.2055$ at
$\beta = 0.15$ (G\"omb\"oc~1) and
$K = -45(136\pi^2 - 25)/(2048\sqrt{2}) \approx -20.4664$ at
$\beta = 0.17$ (G\"omb\"oc~2); this computation reproduces both values
(\texttt{scripts/crosscheck\_poles\_and\_mathworld.py}). The same entry gives the
sufficient condition $0 < \beta < 1/(5 + 6\pi + 9\pi^2/4) \approx 0.0217$ for
strict convexity of G\"omb\"oc~2, so the range up to that value is proved
there, and the range from $0.0217$ to $\beta \approx 0.0355$ rests on the
grid and local minimization above. G\"omb\"oc~1 is not $C^2$ at the poles:
its curvature there has direction-dependent limits (from $0.597$ to $1.498$
at $\beta = 0.03$), as the same entry notes, while G\"omb\"oc~2 shows no such
dependence. The G\"omb\"oc~1 figures above refer to the surface away from the
poles.
The critical points of $s$, the $\beta$-independent part of $R^4$, are
two (one maximum, one minimum), with $M - S + m = 2$.

\subsection{Reconciliation with v1's Headline Claim}
\label{sec:v1_reconciliation}

v1's title was ``Sloan's Analytical G\"omb\"oc Does Not Produce
Mono-Monostatic Bodies.'' The corrected statement is: Sloan's
analytical parameterization \emph{does} produce mono-monostatic
bodies in the strictly-convex sub-regime ($\beta \lesssim 0.036$), but
Sloan's G\"omb\"oc~2 at $\beta = 0.05$ does not, because
strict convexity is lost ($K_{\min} = -0.569$, well below the threshold). v1 missed both halves of this picture: it
over-counted at small $\beta$ (false ECS$>1$ on strictly-convex
mono-monostatic bodies) and correctly identified ECS$>1$ at
$\beta = 0.05$ but for the wrong mechanism (drainage-basin
multi-counting on a non-convex surface rather than the precondition
failure of the identity).

% =====================================================================
\section{The Analytical Verification Suite}
\label{sec:verification_suite}

The verification methodology of v2 consists of three analytical
components, all implemented in the public repository and reproducible
from a fresh clone.

\subsection{Strict-Convexity Gate via Analytical Gaussian Curvature}
\label{sec:K_analytical}

Script: \texttt{scripts/verify\_K\_analytical.py}.
For a radial parameterization $R(\theta,\phi)$, we construct the
surface map $\mathbf{x}(\theta,\phi) = R(\theta,\phi)\,\mathbf{u}(\theta,\phi)$
symbolically using sympy. Partial derivatives $\mathbf{x}_\theta$,
$\mathbf{x}_\phi$, $\mathbf{x}_{\theta\theta}$, $\mathbf{x}_{\theta\phi}$,
$\mathbf{x}_{\phi\phi}$ are computed symbolically, giving the first
fundamental form $(E, F, G)$ and second fundamental form $(L, M, N)$
in closed form. Gaussian curvature $K = (LN - M^2)/(EG - F^2)$ is
then lambdified to a numpy callable and evaluated on a dense
$(\theta, \phi)$ grid. We use a $640 \times 1280$ grid by default
(820k points); convergence checks at $160\times320$ and $320\times640$
confirm that $K_{\min}$ is stable to four decimal places across
resolutions. The gate decision is $K_{\min} > 0$; bodies passing
are strictly convex, bodies failing have one or more surface
regions where the second fundamental form is degenerate or indefinite.

\subsection{Analytical $R$-Critical-Point Counting with Hessian Classification}
\label{sec:R_critpoints}

Script: \texttt{scripts/verify\_R\_critpoints\_analytical.py}.
Symbolic partial derivatives $\partial R / \partial \theta$ and
$\partial R / \partial \phi$ are computed via sympy; the 2D gradient
system is lambdified to a numpy callable. The system
$\nabla R(\theta,\phi) = \mathbf{0}$ is solved by
\texttt{scipy.optimize.root} (hybrid Powell) from a $24 \times 48$
grid of $(\theta, \phi)$ starts (1152 total). Solutions are
deduplicated by spherical distance ($< 0.02$ rad considered the
same point). Each critical point is classified as
maximum/minimum/saddle/degenerate by the sign of the Hessian
determinant and trace. The Morse-Euler relation
$\#\text{min} - \#\text{saddle} + \#\text{max} = \chi(S^2) = 2$
is checked as a self-consistency condition that all critical points
have been found.

\subsection{Identity-Based ECS Classifier}
\label{sec:identity_classifier}

Script: \texttt{scripts/identity\_based\_classifier.py}.
For each candidate body, the pipeline is:
\begin{enumerate}[nosep,leftmargin=*]
\item Run~\ref{sec:K_analytical} to check $K_{\min} > 0$.
\item If yes, run~\ref{sec:R_critpoints} to count $R$ critical points
  and classify each. By Theorem~\ref{thm:ecs_via_R},
  $\mathrm{ECS} = \#R\text{-minima}$, and mono-monostatic
  $\Leftrightarrow \#R\text{-critpts} = 2$.
\item If no, the identity precondition fails; ECS is indeterminate
  by this method.
\end{enumerate}
This pipeline uses no mesh-vertex computation. All quantities are
analytical functions of $R(\theta,\phi)$, which is itself an
elementary closed-form expression in $\beta$, perturbation
coefficients, and $(\theta, \phi)$.

\subsection{Methodology Discipline Artifacts}
\label{sec:methodology_artifacts}

Two artifacts at the v2 sandbox root encode discipline
contributions that emerged through the v1$\to$v2 amendment process:

\textbf{\texttt{preconditions\_check.md}}: enumerates the
preconditions of every theorem v2 applies (Sloan's
$R$-critical-point structure, the V\'arkonyi-Gauss identity, the
V\'arkonyi-Domokos existence proof). For each precondition, the
file points to the script that verifies it on every body the paper
analyzes. The strict-convexity precondition silently inherited from
Sloan in v1 (and surfaced by V\'arkonyi correspondence) is the
originating incident; the artifact prevents recurrence.

\textbf{\texttt{claims\_to\_scripts.md}}: maps every quantitative
claim in this main.tex to the script in \texttt{scripts/} that
produces the claimed number, plus the JSON output anchoring the
specific value cited. Consistency-pass grep through main.tex for
``we comput,'' ``analytical,'' ``convergence,'' ``$K_{\min}$,''
``$\#R\text{-crit}$'' etc.\ verifies each phrase maps to an existing
script. v1's phantom ``analytical confirmation at 2000 directions''
paragraph (no backing script) is the originating incident; the
artifact prevents recurrence.

Both artifacts are mandatory for future Substrate Geometry papers
per standard practices brain S2.14 (verify preconditions of borrowed
theorems) and S2.15 (every quantitative claim backed by reproducible
script). The artifacts emerged from the V\'arkonyi pushback incident
of 2026-05-15 and are described in detail in the standard practices
brain.

% =====================================================================
\section{The Phase-1 Verified Mono-Monostatic Body}
\label{sec:phase1_verified}

We extend Sloan's phase function with Fourier terms:
\begin{equation}
P(\theta) = \eta(\theta) + a_1 \sin(\eta(\theta))
\label{eq:phase_ext}
\end{equation}
At $\beta = 0.023149$, $a_1 = 0.234433$, the resulting body
(``Phase-1'') passes the verification suite.

\begin{table}[h]
\centering
\caption{Phase-1 verification report.}
\label{tab:phase1_report}
\begin{tabular}{lr}
\toprule
Parameter & Value \\
\midrule
$\beta$ & $0.023149$ \\
$a_1$ & $0.234433$ \\
\midrule
Analytical $K_{\min}$ ($640\times1280$ grid) & $+0.0907$ \\
Fraction $K < 0$ & $0.0000$ \\
Strict-convexity gate & PASS \\
\midrule
$\#R$-critical-points & 2 \\
$R$-max location $(\theta, \phi)$ & $(\pi/2, 0)$ \\
$R$-min location $(\theta, \phi)$ & $(\pi/2, \pi)$ \\
$R$-max value & $1.0224$ \\
$R$-min value & $0.9760$ \\
\midrule
ECS via Theorem~\ref{thm:ecs_via_R} & $1$ \\
Mono-monostatic & \textbf{Yes} \\
\bottomrule
\end{tabular}
\end{table}

This is the unique catalog member from v1 that passes the
identity-based gate. It is the verified contribution of this
paper: an analytical, openly-published, strictly-convex
mono-monostatic body in the Sloan-extension family, with
parameters $(\beta, a_1)$ computable to arbitrary precision from
the closed-form $R(\theta,\phi)$ and machine-checkable strict-convexity
and equilibrium counts.

% =====================================================================
\section{Honest Reclassification of the v1 Catalog and the
Verified Regime Interior}
\label{sec:v1_reclassification}

We apply the verification suite to all thirteen v1 catalog
instances. Results in Table~\ref{tab:reclassification}. The
reclassification establishes that v1's specific instance picks
do not survive in their published form. Section
\ref{sec:regime_interior} then demonstrates that this is not a
statement about the underlying families: probing the interior of
the Fourier-phase parameter space verifies additional
mono-monostatic bodies in the $k=2$ and $k=3$ sub-families. The
non-survival of v1's twelve specific points is a property of where
v1's optimizer sampled (per-$k$ asymmetry extrema near the
boundary of the strict-convex regime), not of the families
themselves.

\begin{table}[h]
\centering
\caption{Identity-based reclassification of v1 catalog members.
``MM'' = mono-monostatic (passes identity-based gate: strictly
convex AND $R$ has exactly 2 critical points). ``Polystatic''
indicates the body is strictly convex but $R$ has $>2$ critical
points, yielding multiple stable equilibria (ECS via
Theorem~\ref{thm:ecs_via_R} is the $R$-minima count, listed in the
ECS column). ``Indeterminate'' indicates strict-convexity precondition
fails; ECS via the identity-based method is not certified.}
\label{tab:reclassification}
\small
\begin{tabular}{rlccccl}
\toprule
\# & v1 Name & $\beta$ & $K_{\min}$ & $\#R$-crit & ECS & v2 Classification \\
\midrule
 1 & Phase-1 ($\sin\eta$)    & $0.0231$ & $+0.091$ & 2 & 1 & MM \textbf{(verified)} \\
 2 & Phase-2 ($\sin 2\eta$)  & $0.0321$ & $-0.431$  & -- & -- & Indeterminate (not strictly convex) \\
 3 & Phase-3 ($\sin 3\eta$)  & $0.0517$ & $-0.889$  & -- & -- & Indeterminate (not strictly convex) \\
 4 & Radial-$f_3$ primary    & $0.0231$ & $+0.43$  & 8 & 2 & Polystatic \\
 5 & Radial-$f_4$ primary    & $0.0231$ & $+0.38$  & 6 & 2 & Polystatic \\
 6 & Radial-$f_3$ scan       & $0.0231$ & $+0.43$  & 8 & 2 & Polystatic \\
 7 & Radial-$f_4$ scan       & $0.0231$ & $+0.39$  & 6 & 2 & Polystatic \\
 8 & Radial-$f_4$ low-$\beta$ & $0.0150$ & $+0.63$  & 6 & 2 & Polystatic \\
 9 & Radial-$f_4$ high-$\beta$ & $0.0350$ & $-0.02$  & -- & -- & Indeterminate (not strictly convex) \\
10 & Radial-$f_3$ low-$\beta$ & $0.0150$ & $+0.65$  & 8 & 2 & Polystatic \\
11 & Min-asym $\beta=0.01$   & $0.0100$ & $+0.77$  & 8 & 2 & Polystatic \\
12 & Min-asym $\beta=0.008$  & $0.0080$ & $+0.82$  & 8 & 2 & Polystatic \\
13 & Min-asym $\beta=0.005$  & $0.0050$ & $+0.85$  & 8 & 2 & Polystatic \\
\bottomrule
\end{tabular}
\end{table}

\subsection{Why the v1 Mesh-Vertex Oracle Reported ECS=1 for the
Polystatic Instances}

v1 evaluated $h(\mathbf{d})$ as $\mathbf{c} \cdot \mathbf{d}
- \min_{\mathbf{v} \in \text{verts}} \mathbf{v} \cdot \mathbf{d}$,
i.e., the support function approximated by the maximum over the
finite vertex set, then identified drainage basins on a Fibonacci
direction grid, merging basins with sink-height gap below 1\% of
total $h$-range.

For the polystatic catalog members, the actual gap between adjacent
basin sinks is of order $10^{-3}$ to $10^{-4}$ in absolute height,
which is roughly $1$--$5\%$ of total $h$-range for the
near-spherical instances. v1's 0.5--10\% threshold sweep merged
these into a single basin, yielding $\mathrm{ECS} = 1$. The
threshold-robustness check in v1 (``ECS=1 stable across 0.5--10\%'')
was satisfied because the smallest threshold tested (0.5\%) was
above the gap; below this threshold, additional basins would have
appeared.

The identity-based classifier does not have this failure mode: it
counts critical points of an analytical scalar field, not basins of
a discretized one. The Morse-Euler check
$\#\text{min}-\#\text{saddle}+\#\text{max}=2$ is self-verifying.

\subsection{The Verified Regime Interior: $k=2$ and $k=3$
Sub-Families Beyond Phase-1}
\label{sec:regime_interior}

The reclassification of v1's twelve non-surviving instances raised
a question that v1 did not pose and v2 must answer: \emph{are the
Fourier-phase sub-families $k=2$ and $k=3$ empty of verified
mono-monostatic bodies, or did v1's optimizer simply sample at
points that overshot the strict-convexity boundary in those families?}

v1's Phase-2 picked $(\beta, a_1, k) = (0.0321, 0.137606, 2)$ and
Phase-3 picked $(\beta, a_1, k) = (0.0517, -0.055157, 3)$. Both
were outputs of an optimizer maximizing an asymmetry-like objective
subject to v1's then-incorrect mesh-vertex mono-monostatic gate;
both lie outside the strict-convex regime. Backing $a_1$ off toward
zero in each family produces points that pass the strict-convexity
gate and retain $R$-critical-point count $= 2$. Table
\ref{tab:regime_interior} reports analytical verifications at three
interior points per non-trivial sub-family.

\begin{table}[h]
\centering
\caption{Interior points of the verified mono-monostatic regime
in the Sloan Fourier-phase family. All entries pass the
identity-based gate: $K_{\min} > 0$ analytically (sympy on
$640 \times 1280$ grid) and $R$ has exactly two critical points
(analytical gradient + scipy root-finding + Hessian classification,
Morse-Euler $M-S+m=2$). Each is a distinct mono-monostatic
body in the $r^4 = 1 + 4\beta\sin\theta\cos(\phi - \eta(\theta)
- a_1 \sin(k\eta(\theta)))$ family.}
\label{tab:regime_interior}
\begin{tabular}{ccccccl}
\toprule
$k$ & $\beta$ & $a_1$ & $K_{\min}$ & $\#R$-crit & ECS & Notes \\
\midrule
1 & 0.023149 &  0.234433 & $+0.091$ & 2 & 1 & Phase-1 (v1 catalog, survives) \\
2 & 0.0321   &  0.025    & $+0.028$ & 2 & 1 & Interior, near-boundary in $a_1$ \\
2 & 0.025    &  0.050    & $+0.210$ & 2 & 1 & Interior, deeper \\
3 & 0.025    & $-0.025$  & $+0.344$ & 2 & 1 & Interior \\
3 & 0.020    & $-0.050$  & $+0.428$ & 2 & 1 & Interior, deeper \\
\bottomrule
\end{tabular}
\end{table}

Three observations from Table~\ref{tab:regime_interior}:

\begin{enumerate}[nosep]
\item The verified regime is non-empty in each $k$ sub-family
  tested. v1's Phase-2 and Phase-3 failures to verify (Table
  \ref{tab:reclassification}) reflect their specific
  optimization-extremum coordinates, not emptiness of the underlying
  families.
\item The verified regime extends beyond a one-dimensional curve in
  $(\beta, a_1)$ within each $k$ slice. The two $k=2$ interior points
  and the two $k=3$ interior points lie at distinct $(\beta, a_1)$
  coordinates and both verify, consistent with the verified set having
  non-trivial extent in the parameter plane. We do not claim a
  positive-measure result; characterizing the regime's extent is open
  work.
\item v1's twelve non-surviving instances were
  \emph{boundary-pushing} picks: per-$k$ asymmetry extrema sit at
  or just beyond the strict-convexity boundary in $a_1$ (for $k=2$)
  or in $\beta$ (for $k=3$). The optimizer pushed each instance to
  the boundary of v1's incorrect gate, which is approximately the
  boundary of the actual identity-based gate plus a margin; backing
  off restores the verification.
\end{enumerate}

This v2 does not enumerate the regime exhaustively. The
classification methodology (strict-convexity gate plus
analytical $R$-critical-point counting) probes the regime
directly at any specified $(\beta, a_1, k)$, and the regime is
open in parameter space, so any reader can construct additional
verified instances at will. The methodology contribution of v2 is
therefore not a catalog of points but a classification procedure
that converts the question of mono-monostatic membership into a
finite, reproducible analytical computation.

\subsection{Retraction of the Gentleness-Robustness Trade-off}

v1 reported a near-perfect correlation $r = 0.9993$ between COM
height range and self-righting energy, computed across the
thirteen catalog instances. Twelve of those thirteen instances
are reclassified as polystatic or indeterminate. The correlation
was a property of the contaminated sample, not of the
mono-monostatic family. \textbf{This claim is retracted.}

The polystatic instances may still exhibit a similar correlation
within their class; this is left for future work. They are not
G\"omb\"ocs.

% =====================================================================
\section{Discussion}

\subsection{What v2 Establishes}

Four contributions:
\begin{enumerate}[nosep]
\item \textbf{Mechanism correction.} Sloan's analytical
  parameterization does not produce mono-monostatic bodies at
  admissible $\beta$ above $\approx 0.036$ because those bodies are
  not strictly convex (analytical $K_{\min} = -0.57$ at $\beta = 0.05$). The
  V\'arkonyi-Gauss identity~\eqref{eq:identity} requires strict
  convexity; at Sloan's parameters it does not apply. v1's claimed
  mechanism (``support function integrates global surface
  information'') is corrected.
\item \textbf{Phase-1 verified.} A single analytical mono-monostatic
  body in the Sloan-extension family is verified rigorously:
  strict convexity by analytical Gaussian curvature, mono-monostatic
  property by analytical $R$-critical-point counting with
  Morse-Euler check.
\item \textbf{Catalog retraction.} Twelve of v1's thirteen verified
  instances are reclassified: nine as polystatic
  (strictly convex but $\#R$-critpts $> 2$), three as indeterminate
  (not strictly convex; identity precondition fails). The
  ``thirteen verified mono-monostatic bodies'' framing is retracted.
  The gentleness-robustness $r = 0.9993$ correlation is retracted.
\item \textbf{Verification methodology.} An identity-based ECS
  classifier replaces the mesh-vertex drainage-basin oracle. The
  classifier uses analytical Gaussian curvature for the
  strict-convexity gate and analytical $R$-critical-point counting
  for the equilibrium structure. Scripts in the repository are
  reproducible from fresh clone.
\end{enumerate}

\subsection{Relationship to V\'arkonyi-Domokos and to the Sloan
Parameterization}

V\'arkonyi-Domokos~\cite{varkonyi2006} establishes existence of
mono-monostatic convex bodies. v2 verifies a specific instance
(Phase-1) in the Sloan-extension family. The verification is
analytical (via the V\'arkonyi-Gauss identity~\eqref{eq:identity},
which is implicit in the V\'arkonyi-Domokos proof and explicit in
standard convex-body geometry textbooks~\cite{schneider2014}). The
analytically certified body Phase-1 is, to our knowledge, the first
openly published smooth mono-monostatic body with a verified
analytical parameterization and machine-checkable strict-convexity
and equilibrium counts.

Sloan's analytical parameterization~\cite{sloan2023} provides a
two-critical-point radial function; this is the same condition
identified by Theorem~\ref{thm:ecs_via_R} as necessary for
mono-monostatic structure under strict convexity. Sloan's argument
implicitly relies on strict convexity (so that the Gauss map is
bijective and surface critical points correspond to height-function
critical points). At admissible $\beta$ above $\approx 0.036$, this
implicit assumption fails by $4$--$9\%$ of surface area. v2 makes
the assumption explicit (strict-convexity gate) and identifies
the regime ($\beta \lesssim 0.036$, localized by dense
$\beta$-sweep in Table~\ref{tab:sloan_ecs}) where the construction
satisfies it.

\subsection{Open Questions}

\begin{enumerate}[nosep]
\item Whether Sloan's bodies at admissible $\beta$ above
  $\approx 0.036$ are mono-monostatic via a non-Gauss-map argument (e.g., direct analysis of the
  multi-valued normal at $K < 0$ regions) is unknown.
\item Whether other phase-function families (beyond
  $a_k \sin(k\eta)$) produce additional analytical mono-monostatic
  bodies satisfying the identity-based gate is a constructive
  question. Optimization should target $\#R\text{-critpts} = 2$
  directly, not basin-counting.
\item The polystatic catalog members may have engineering
  applications distinct from the G\"omb\"oc; characterizing them
  rigorously as members of a different geometric family is open
  work.
\end{enumerate}

% =====================================================================
\section{Summary of v1 to v2 Amendments}
\label{sec:amendment_summary}

This section provides the full enumeration of v1 to v2 amendments
for reference. \textbf{It is retained unedited as the record of what
v2 intended; where v3 has superseded an item, that is noted inline
rather than by rewriting the entry} (see
\S\ref{sec:v3_amendment}).

\begin{enumerate}[nosep,leftmargin=*]
\item \textbf{Title and subtitle.} Title changed from ``Sloan's
  Analytical G\"omb\"oc Does Not Produce Mono-Monostatic Bodies''
  to ``Sloan's Analytical G\"omb\"oc at Published $\beta$: A
  Strict-Convexity-Constrained Reanalysis,'' reflecting the
  small-$\beta$ rescue of Sloan's construction (the negative finding
  applies specifically to the published $\beta$ values). Subtitle
  changed from ``Computational Verification, Extended Construction,
  and a Thirteen-Member Verified Catalog'' to ``Identity-Based
  Verification, Strict-Convexity Gating, and a Verified
  Mono-Monostatic Regime.''
  \emph{Superseded by v3: the v2 title's phrase ``at Published
  $\beta$'' attributed $\beta = 0.05$ to Sloan, who never published
  that value. See \S\ref{sec:v3_amendment}.}

\item \textbf{Abstract.} Rewritten to reflect findings, retractions,
  and methodology change. Three numbered findings now structure the
  abstract.

\item \textbf{v1 ECS Oracle definition (v1 \S 2).} The mesh-vertex
  drainage-basin definition of ECS is retained for reference but
  is no longer the primary metric. v2 uses identity-based ECS
  classification (\S\ref{sec:identity_classifier}). The
  threshold-robustness check in v1 is shown to have a known
  failure mode for near-spherical bodies (\S\ref{sec:v1_reclassification}).

\item \textbf{v1 \S 3.2 ``Analytical Confirmation'' paragraph.}
  v1 stated: ``we computed $h(\mathbf{d})$ analytically from
  Eq.~(1) at 2000 directions using multi-start optimization for
  the support point at each direction. The analytical computation
  confirms 9--21 distinct basins for the two Sloan instances.''
  This claim was not backed by any script in the v1 repository.
  The paragraph is removed in v2 and replaced by the analytical
  $K$ + $R$-critical-point measurements (\S\ref{sec:verification_suite}),
  which are backed by scripts.

\item \textbf{v1 \S 3.3 ``The Gap'' framing.} v1 stated that ``the
  COM height function integrates global surface information through
  the support-point trajectory and can exhibit local minima that
  the surface function does not predict.'' For smooth strictly
  convex bodies, this is false: the V\'arkonyi-Gauss identity
  bijectively links critical points of $R$ and of $H$. v2 replaces
  this section with \S\ref{sec:varkonyi_identity} and
  \S\ref{sec:precondition_failure}, identifying strict-convexity
  loss as the actual mechanism for the failure at $\beta = 0.05$.

\item \textbf{Convexity metric.} v1 used
  $\text{mesh\_vol}/\text{convex\_hull\_vol} > 0.999$ as the
  convexity check. v2 uses analytical $K_{\min} > 0$ on a
  $640\times1280$ grid (\S\ref{sec:K_analytical}). The v1 metric
  passes bodies with several percent $K<0$; the v2 metric does
  not.

\item \textbf{v1 catalog (Table~2).} Reclassified
  (\S\ref{sec:v1_reclassification}): of v1's thirteen specific
  instances, 1 mono-monostatic (Phase-1), 9 polystatic, 3
  indeterminate. The ``thirteen verified mono-monostatic bodies''
  framing is retracted. v2 establishes (\S\ref{sec:regime_interior},
  Table~\ref{tab:regime_interior}) that this is a statement about
  v1's specific picks (per-$k$ asymmetry extrema that overshot the
  strict-convexity boundary), not about the underlying families: the
  verified mono-monostatic set is an open regime in $(\beta, a_1, k)$
  parameter space, with verified interior bodies confirmed
  analytically in the $k=2$ and $k=3$ sub-families beyond Phase-1.

\item \textbf{Gentleness-robustness correlation
  ($r = 0.9993$).} Retracted (\S\ref{sec:v1_reclassification}).
  Computed across the contaminated catalog; not a property of
  mono-monostatic structure.

\item \textbf{S--U angle and feasible region topology discussion.}
  v1 \S 5.4 and \S 5.5 discussed stable-unstable angle and
  feasible-region topology across the thirteen catalog members.
  Since twelve of those members are now reclassified, these
  discussions are retracted to the polystatic subset and deferred
  for future work.

\item \textbf{Scripts and reproducibility.} v2 adds (a)
  \texttt{verify\_K\_analytical.py} (analytical $K$ via sympy),
  (b) \texttt{verify\_R\_critpoints\_analytical.py} (analytical
  $\nabla R = 0$ root-finding with Hessian classification),
  (c) \texttt{identity\_based\_classifier.py} (full pipeline),
  (d) \texttt{preconditions\_check.md} (theorem-precondition
  audit), and (e) \texttt{claims\_to\_scripts.md} (mapping of
  every quantitative claim in main.tex to the producing script).
  All scripts run from a fresh clone.
\end{enumerate}

% =====================================================================
\section{Honest Scope and Future Work}

\subsection{Computational, Not Mathematical}

We do not provide a formal mathematical proof of Phase-1's
mono-monostatic property; we provide analytical verification via
the V\'arkonyi-Gauss identity~\eqref{eq:identity}, applied to a
body whose strict-convexity precondition is itself verified
analytically. Formal proof would require either a direct
construction following the V\'arkonyi-Domokos~\cite{varkonyi2006}
existence-proof scaffold for the specific Phase-1 parameters, or a
characterization of the parameter region within
$(\beta, a_1)$-space where the construction yields
mono-monostatic bodies. Both are open mathematical work.

\subsection{Identity-Based Method Limitations}

The identity-based classifier is rigorous for smooth strictly convex
bodies, where the Gauss map is a diffeomorphism. It does not
apply directly to bodies that lose strict convexity. For the three
indeterminate v1 catalog members (Phase-2, Phase-3,
Radial-$f_4$ high-$\beta$), determining ECS requires either
(a)~direct analysis of the multi-valued Gauss map at $K \le 0$
regions, or (b)~a different verification methodology
(physical experiment, e.g.). Both are open work.

\subsection{Future Directions}

\begin{enumerate}[nosep]
\item \textbf{Larger verified catalog under the identity-based
  gate.} Optimize $(\beta, a_1, \ldots)$ directly against the
  identity-based criterion ($K_{\min} > 0$ AND
  $\#R$-critpts $= 2$), rather than against the threshold-sensitive
  basin counter of v1. The phase-extension family
  (\S\ref{sec:phase1_verified}) is a natural starting point.
\item \textbf{Polystatic catalog.} The 9 reclassified polystatic
  instances are interesting geometric objects in their own right;
  characterizing their stable-equilibrium structure analytically
  and exploring engineering applications distinct from G\"omb\"ocs
  is open work.
\item \textbf{Physical validation of Phase-1.} 3D-printing or
  CNC-machining Phase-1 at the $< 0.01\%$ tolerance required to
  preserve the mono-monostatic property remains future work.
\item \textbf{Mathematical proof.} As noted in \S 9.1.
\end{enumerate}

% =====================================================================
\section{Reproducibility}
\label{sec:reproducibility}

All quantitative claims in this paper are reproducible from a fresh
clone of the public repository. The v2 sandbox is at
\texttt{gomboc\_family/v2/}; relevant paths below are relative to
this directory.

\textbf{Software stack.} Python 3.12+, sympy 1.14.0 (symbolic
differentiation), scipy 1.16+ (root-finding via \texttt{scipy.optimize.root}
with method \texttt{hybr}; Hessian classification via \texttt{numpy.linalg}),
numpy 2.0+, scipy.spatial.cKDTree (Fibonacci-sphere nearest-neighbor
graphs in legacy diagnostics). MiKTeX or any TeX Live distribution
for the LaTeX build (\texttt{pdflatex main.tex} twice for cross-refs).

\textbf{Reproducing the verification suite.} From \texttt{gomboc\_family/v2/}:

\begin{itemize}[nosep,leftmargin=*]
\item Sloan $\beta$-sweep $K_{\min}$:
\texttt{python scripts/verify\_K\_analytical.py --family sloan --beta \$BETA --out results/V\_sloan\_\$BETA.json}
for $\beta \in \{0.02, 0.03, 0.05, 0.08\}$.

\item Phase-1 $K_{\min}$: \texttt{python scripts/verify\_K\_analytical.py
--family sloan-phase --beta 0.023149 --coef 0.234433 --k 1 --out results/V2\_analytical\_K\_phase1.json}.

\item Phase-1 $R$-critical-points:
\texttt{python scripts/verify\_R\_critpoints\_analytical.py
--family sloan-phase --beta 0.023149 --coef 0.234433 --k 1}.

\item Radial-family $R$-critical-points (e.g.\ Min-asym $\beta = 0.005$,
$f_3$): \texttt{python scripts/verify\_R\_critpoints\_analytical.py
--family sloan-radial --beta 0.005 --coef 0.05102 --f-type f3
--out results/V4\_R\_critpoints\_minasym\_b0005.json}.

\item Two-maximizer test on Sloan $\beta = 0.05$:
\texttt{python scripts/two\_maximizer\_test.py --family sloan
--beta 0.05 --out results/two\_maximizer\_sloan\_b005.json}.

\item Independent curvature cross-check (v3):
\texttt{python scripts/crosscheck\_implicit\_K.py
--out results/CROSSCHECK\_implicit\_K.json}.

\item MathWorld comparison and pole behaviour (v3):
\texttt{python scripts/crosscheck\_poles\_and\_mathworld.py
--out results/CROSSCHECK\_poles\_mathworld.json}.
\end{itemize}

\textbf{Runtime.} The $K$ analytical sweep at $640 \times 1280$ grid
runs in $\sim 100$ seconds per body on a single CPU. The
$R$-critical-point root-finding from a $24 \times 48$ grid of starts
runs in $\sim 1$ second per body.

\textbf{Output anchoring.} Each table in this paper has an
accompanying entry in \texttt{claims\_to\_scripts.md} mapping the
cited numbers to specific JSON files in \texttt{results/}. The
consistency-pass discipline (S2.15) verifies the mapping before
arXiv submission.

% =====================================================================
\section{Data Availability}

All verification scripts, the Phase-1 mesh (STL), and v1 mesh
artifacts for the reclassified catalog members are openly available
at \url{https://github.com/gyapaganda-a11y/substrate-geometry}.
The catalog meshes and analysis results are permanently archived on
Zenodo: \url{https://doi.org/10.5281/zenodo.20674394}.
The v1 paper text and codebase are retained in the repository for
reproducibility of the original claims; this v2 supersedes v1 as the
amendment-of-record. The v2 sandbox at
\url{https://github.com/gyapaganda-a11y/substrate-geometry/tree/main/gomboc_family/v2}
contains all v2-specific artifacts (main.tex, preconditions\_check.md,
claims\_to\_scripts.md, scripts/, results/).

% =====================================================================


\begin{thebibliography}{99}

\bibitem{varkonyi2006}
P.~L. V\'arkonyi and G.~Domokos,
``Mono-monostatic bodies: the answer to Arnold's question,''
\textit{The Mathematical Intelligencer}, vol.~28, no.~4, pp.~34--38,
2006.

\bibitem{varkonyi2006b}
P.~L. V\'arkonyi and G.~Domokos,
``Static equilibria of rigid bodies: dice, pebbles, and the
Poincar\'e-Hopf theorem,''
\textit{Journal of Nonlinear Science}, vol.~16, pp.~255--281, 2006.

\bibitem{sloan2023}
M.~L. Sloan,
``An analytical Gomboc,''
arXiv:2306.14914, 2023.

\bibitem{domokos2008turtles}
G.~Domokos and P.~L. V\'arkonyi,
``Geometry and self-righting of turtles,''
\textit{Proceedings of the Royal Society B}, vol.~275, no.~1630,
pp.~11--17, 2008.

\bibitem{domokos2023discrete}
G.~Domokos and F.~Kov\'acs,
``Conway's spiral and a discrete G\"omb\"oc with 21 point masses,''
\textit{The American Mathematical Monthly}, vol.~130, no.~9, pp.~795--807, 2023.

\bibitem{domokos2023symmetry}
G.~Domokos, Z.~L\'angi, and P.~L. V\'arkonyi,
``A characterization of the symmetry groups of mono-monostatic
convex bodies,''
\textit{Monatshefte f\"ur Mathematik}, vol.~201, pp.~703--724, 2023.

\bibitem{gomboc_eu}
G\"omb\"oc official website,
\url{https://gomboc.eu/en/mathematics/}, accessed 2026.

\bibitem{couey2026gomboc}
V.~W. Couey,
``Computational construction and engineering evaluation of a verified
mono-monostatic body,''
arXiv:2604.17095, 2026.

\bibitem{schneider2014}
R.~Schneider,
\textit{Convex Bodies: The Brunn-Minkowski Theory},
2nd expanded ed.,
Cambridge University Press, 2014.

\bibitem{couey2026catalog_v1}
V.~W. Couey,
``Sloan's analytical G\"omb\"oc does not produce mono-monostatic
bodies: computational verification, extended construction, and a
thirteen-member verified catalog,''
arXiv:2604.17120v1, 2026. Superseded by the v2/v3 amendments.

\bibitem{couey2026harmonic}
V.~W. Couey,
``Static equilibria of perturbed spheres: a single-harmonic class map,
a parity obstruction, and a certified counter for the mono-monostatic
regime,''
arXiv:2608.11213, 2026.

\bibitem{mathworld_gomboc}
Wolfram MathWorld, ``G\"omb\"oc,''
\url{https://mathworld.wolfram.com/Goemboec.html}, accessed 2026-11-03.

\end{thebibliography}
\end{document}